\documentclass[12pt]{article}
\usepackage{amssymb,graphicx}
\usepackage{amsmath,amsfonts}
\usepackage{epsfig}

\def\l{\left(}
\def\r{\right)}

\newcommand{\be}{\begin{equation}}
\newcommand{\ee}{\end{equation}}
\newcommand{\bea}{\begin{eqnarray}}
\newcommand{\eea}{\end{eqnarray}}
\newcommand{\bg}{\begin{gather}}
\newcommand{\eg}{\end{gather}}
\newcommand{\bseq}{\begin{subequations}}
\newcommand{\eseq}{\end{subequations}}

\newcommand\lp{\left(}
\newcommand\rp{\right)}
\newcommand\Ly{\mbox{Lyman-$\alpha$} }
\newcommand\mpl{M_\text{Pl}}

\begin{document}
\begin{flushright}
%Prepint Number
\end{flushright}
\vspace{10pt}
\begin{center}
  {\LARGE \bf Constraining sterile neutrino dark matter by
  phase-space density observations} \\
\vspace{20pt}
%\medskip
D.~Gorbunov, A.~Khmelnitsky and V.~Rubakov\\
\vspace{15pt}
\textit{
Institute for Nuclear Research of
         the Russian Academy of Sciences,\\  60th October Anniversary
  Prospect, 7a, 117312 Moscow, Russia}\\

    \end{center}
    \vspace{5pt}

\begin{abstract}

We apply phase-space density considerations to obtain lower bounds on
the mass of sterile neutrino as dark matter candidate. The bounds are
different for non-resonant production, resonant production in the
presence of lepton asymmetry and production in decays of heavier
particles. In the former case our bound is comparable to, but
independent of the \Ly bound, and together with X-ray upper limit it
disfavors non-resonantly produced sterile neutrino dark matter. An
interesting feature of the latter case is that \emph{warm} dark matter
may be composed of \emph{heavy} particles.

\end{abstract}
\vspace{5pt}
%%%%%%%%%%%%%%%%%%%%%%%%%%%%%%%

One candidate for dark matter particle is sterile neutrino. In the
first place, the existence of sterile neutrinos is favored by the
discovery of neutrino oscillations. Furthermore, by adding sterile
neutrinos to the Standard Model and fine-tuning parameters in the
neutrino sector, one is able to explain all known facts in high energy
physics and cosmology without introducing any other new fields except
for inflaton~\cite{nuMSM,Asaka:2005pn,Tkachev}.

The sterile neutrino dark matter should satisfy cosmological and
astrophysical constraints. These constraints depend on the way sterile
neutrinos are produced in the early Universe. There are several
mechanisms capable of generating sterile neutrino in right
amount. These include the non-resonant production via active-sterile
neutrino mixing~\cite{DW}, resonant production in the presence of
lepton asymmetry~\cite{SF,AS-SF}, production in decays of heavier 
particles (see, e.g., Refs.~\cite{Tkachev, decays} in the context of
$\nu$MSM model) and production in scattering (see e.g.,
Ref.~\cite{Khalil}).  
 In this paper we consider astrophysical constraints
on sterile neutrino dark matter generated by the non-resonant
production mechanism. We comment on other production mechanisms
towards the end of this paper.

Sterile neutrino is coupled to active ones through mixing only. Thus,
the mixing angle controls the decay rate of sterile neutrino into
active neutrino and photons. These decays in galactic halos would
produce X-ray emission lines, which are severely
constrained~\cite{Xray} by current observations. Within the
non-resonant production mechanism, the mixing angle and the mass $m$
of sterile neutrino are related to each other by the requirement that
sterile neutrinos make all of the dark matter. So, the X-ray
observations put the upper bound on the mass, $m <
4\,\text{keV}$~\cite{Xray4}.

Sterile neutrino with the mass in the keV range is \emph{warm} dark
matter candidate.  In fact, warm dark matter may be even more
attractive than cold dark matter~\cite{WDMcure}, as it might be able
to address emerging problems of the standard CDM cosmology on small
scales, such as missing satellites~\cite{satellites}, galactic density
profiles~\cite{cores} and angular momentum of spiral
galaxies~\cite{angular}.  However, like any warm dark matter candidate
sterile neutrino should not be too ''warm'' in order to form observed
small scale structures. This implies that its mass should not be too
small.

Well known lower limits on sterile neutrino mass come from the
observations of \Ly\ forest --- multiple absorption lines in spectra of
distant quasars~\cite{LyA, VHS}. Most of these lines correspond to
smooth overdense regions of warm ionized intergalactic medium which
are believed to trace dark matter clustering. To derive the sterile
neutrino mass bounds, one compares the statistical properties of \Ly
absorption lines with those predicted within various cosmological
models by semi-analytical methods and numerical simulations. Recent
\Ly lower bounds on the mass of sterile neutrino dark matter produced
via the   non-resonant production mechanism are $m > 5.6 - 28\,\text{keV}$
depending on data set~\cite{VHS}. Together with the X-ray upper bound
they disfavor the non-resonantly produced sterile neutrino as the dark matter
candidate.

Given the uncertainties in the \Ly constraints, it is worth deriving a
lower bound on sterile
neutrino mass  
in an independent way. In this paper we employ the
Tremaine--Gunn approach, which was successfully used nearly thirty years ago to
rule out the scenario of active neutrino hot dark
matter~\cite{TG}. This approach relies on the fact that coarse-grained
distribution function of collisionless particles decreases in the course of
evolution.

The process of gravitational formation of compact objects from collisionless
dark matter particles is described by the Vlasov equations governing the
evolution of distribution function in self-consistent gravitation
field, see, e.g. Ref.~\cite{galdyn} and references therein. 
One of the most important phenomena taking place in such
systems is mixing of distribution function in phase-space. The phase-space
volume occupied by the system is invariant under Vlasov equation
dynamics. Nevertheless, particle trajectories chaotically disperse, exploring
new areas of phase-space and forming there ever more complicated fine
structure. This results in overall decrease of coarse-grained distribution
function. In particular, values of coarse-grained distribution function cannot
exceed the maximum value of the primordial distribution
function~\cite{violrelax}. The latter property enables one to obtain
constraints on dark matter.

Primordial distribution function of sterile neutrinos produced via non-resonant
oscillations can be approximated by scaled distribution of active neutrinos ---
ultra-relativistic Fermi-Dirac distribution~\cite{DW}, \be\label{eq:df} f(p)=
\frac{g}{(2\pi)^3}\, \frac\beta{e^{p/T_\nu} + 1} \;, \ee where $g = 2$ is the
number of sterile neutrino spin states. The coefficient $\beta$ here is
proportional to the active-sterile mixing angle squared and can be related to
the sterile neutrino mass by demanding that sterile neutrinos constitute all
dark matter in the Universe\footnote{We do not consider here the possibility
that the non-resonant mechanism generates not all, but only a fraction of dark
matter~\cite{Palazzo}.}:
$$
\beta = \l\frac{\Omega_\text{DM}}{0.2}\r  \l\frac{10\,\text{eV}}{m} \r \;.
$$ 
Thus, the maximum value of the primordial distribution function in
this scenario is given by 
\be\label{eq:maxf} \max f(p) =
\frac1{(2\pi)^3}\lp\frac{\Omega_\text{DM}}{0.2}\rp
\lp\frac{10\,\text{eV}}{m}\rp \;.  
\ee

Observationally, the value of the coarse-grained distribution function 
in a galactic halo
can be
estimated by the phase-space density, the ratio between the mass density and
cube of the one-dimensional velocity dispersion in a given volume, $Q \equiv
\rho/\sigma^3$~\cite{Hogan}. In practice, one measures the velocity dispersion
of stars and assumes that it coincides with that of dark matter particles. For
non-relativistic dark matter particles one has
\[
Q = m^4\cdot \frac{n}{\langle \frac13 p^2\rangle^{3/2}} \; ,
\]
where $n$ is their average number density in a halo.  Assuming that the
coarse-grained distribution of halo particles is isotropic,
$f_{halo}(\mathbf{p}, \mathbf{r}) = f_{halo}(p, r)$, one estimates
\be\label{eq:fest}
\frac{n}{\langle p^2\rangle^{3/2}} = \frac{\left[\int{f_{halo}(\mathbf{p},
\mathbf{r}) d^3\mathbf{p}}\right]^{5/2}} {\left[\int{f_{halo}(\mathbf{p},
\mathbf{r}) \mathbf{p}^2 d^3\mathbf{p}}\right]^{3/2}} \sim f_{halo} (p_*, r) \;,
\ee
where $p_*$ is a typical momentum of the dark matter particles.\footnote{In the
  case when the width of the momentum distribution around $p_*$ is small,
  $\Delta p < p_*$, the estimate~\eqref{eq:fest} reads $n/\langle
  p^2\rangle^{3/2} \sim f_{halo} \cdot (\Delta p/ p_*)$. Then, instead
  of~\eqref{eq:Qf}, one has an inequality, $ f_{halo} > Q/(3^{3/2} m^4)$. This
  makes the constraint~\eqref{eq:constraint} even stronger.}  In this way the
  value of the coarse-grained distribution function in a galactic halo is
  estimated as
\begin{equation}
\label{eq:Qf}
  f_{halo} \approx \frac Q {3^{3/2}\, m^4} \; .
\end{equation}
Hence, one arrives at the following constraint on dark matter,
\be\label{eq:constraint}
 \frac Q {3^{3/2}\, m^4} < \max f(p) \; . 
 \ee

The strongest constraints on warm dark matter scenario are obtained by
making use of the highest values of the phase-space density in dark
matter dominated objects. These are found in dwarf spheroidal
satellite galaxies~\cite{Mateo, Dalcanton}. dSph's are the most dark
matter dominated compact objects observed so far, and are conjectured to be
hosted by the smallest possible dark matter halos~\cite{Mateo}. In
recently discovered objects Coma Berenices, Leo IV and Canes
Venaciti II, the value of $Q$ ranges from $5\cdot10^{-3}\,
\frac{M_{\odot}/\text{pc}^3}{\lp\text{km}/\text{s}\rp^3}$ to
$2\cdot10^{-2}\,
\frac{M_{\odot}/\text{pc}^3}{\lp\text{km}/\text{s}\rp^3}$~\cite{Simon}.
We use the first, slightly more conservative value, 
\begin{equation}
\label{not-required}
 Q \equiv q \ \frac{M_{\odot}/\text{pc}^3}{\lp\text{km}/\text{s}\rp^3}=
5\cdot10^{-3}\,
\frac{M_{\odot}/\text{pc}^3}{\lp\text{km}/\text{s}\rp^3} \; .
\end{equation}

Making use of Eqs.~(\ref{eq:maxf}) and~(\ref{eq:constraint}) we obtain the
constraint on the sterile neutrino mass,
\be
\label{eq:result}
 m > 5.7\ \text{keV} \lp\frac{0.2}{\Omega_\text{DM}}\rp^{1/3}
\l \frac{q}{5\cdot10^{-3}} \r^{1/3}
\; .
\ee

Despite the fact that the value of observable $Q$ may not be exactly
the same as the value of the coarse-grained distribution function of
dark matter particles, we consider the bound~(\ref{eq:result}) as
conservative. Indeed, the coarse grained distribution function is
likely to decrease considerably during the non-linear stage of
evolution. For example, some numerical simulations of halo formation
show the decrease of $Q$ by a factor of $10^2 - 10^3$ from input
initial values~\cite{Peirani}. Therefore, further improvement of
understanding of how compact objects are formed by warm dark matter
particles is likely to strengthen the bound~(\ref{eq:result}). We
conclude that the Tremaine--Gunn approach gives another argument,
independent of \Ly, that disfavors the sterile neutrino dark matter
generated by the non-resonant production mechanism.

However, sterile neutrino may still be a dark matter candidate. There
are other generation mechanisms of sterile neutrino dark matter, such
as resonant production in the presence of lepton
asymmetry~\cite{SF,AS-SF}, production in scattering~\cite{Khalil} and
production in decays~\cite{Tkachev,decays}. In these cases the sterile
neutrino decay rate is not directly related to their abundance in the
Universe, so the X-ray mass bound does not apply.

Although the primordial distribution function of sterile neutrinos in
these scenarios differs from~(\ref{eq:df}), lower bounds from both \Ly
and phase-space observations are not expected to change
dramatically. For example, the results presented in Ref.~\cite{AS-SF}
indicate that the maximum value of the distribution function of 3~keV
sterile neutrinos generated by the resonant production in the presence
of lepton asymmetry is about $0.03 - 0.3$ of the maximum of thermal
distribution, $1/(2\pi)^3$ (accounting for 2 spin states), depending
on the value of the lepton asymmetry. In that case one has
\[
 3^{3/2} m^4 \ \max f (p) = (6.5 \cdot 10^{-3} - 6.5 \cdot 10^{-2})\ 
\frac{M_{\odot}/\text{pc}^3}{\lp\text{km}/\text{s}\rp^3} \; ,
\]
at the edge of the constraint (\ref{eq:constraint}), with $Q$ estimated as in
\eqref{not-required}. In this case the correct dark matter density, $\Omega_\nu
\approx 0.2,$ is obtained for the mixing angle in the range $\sin^2(2 \theta)
= 10^{-11} - 10^{-8}$ and even smaller, again depending on the lepton
asymmetry, and the X-ray bound can be satisfied~\cite{AS-SF}.

If sterile neutrinos are produced in scattering processes and do not
equilibrate, as is the case for the production mechanism of
Ref.~\cite{Khalil}, then the 
neutrino distribution is approximately given by~\eqref{eq:df}
with the same coefficient $\beta$ and effective temperature
$T_\nu$. Hence, the maximum of the neutrino distribution function is
the same, as in the case of non-resonant production. So, the
limit~\eqref{eq:result} is applicable for sterile neutrino produced in
scattering as well.

Before proceeding to sterile neutrino production in decays, let us recall that
there is a simple bound on their mass based on Pauli blocking. Namely, the
primordial distribution function cannot exceed the value $2/(2\pi)^3$, so for
distributions saturating this bound, the constraint \eqref{eq:constraint}
translates into \be m > 1.0~\mbox{keV} \l \frac{q}{5\cdot 10^{-3}} \r^{1/4}\; ,
\label{aug14-2}
\ee where we again used the estimate \eqref{not-required}.  This is a
model-independent lower limit on the mass of a fermionic dark matter candidate
(assuming two spin states).  In what follows we do not take into account Pauli
blocking, with understanding that if the limits obtained are weaker than
\eqref{aug14-2}, then the limit \eqref{aug14-2} applies.

Sterile neutrinos may be produced in decays of relativistic thermalized
particles\footnote{We assume here for simplicity that these particles interact
sufficiently weakly with the rest of the cosmic plasma, so that sterile neutrino
production in scattering processes is negligible. In the opposite case the mass
bound is similar to \eqref{eq:result}.} of mass $M$ and partial decay width at
rest into sterile neutrinos $\Gamma$. In that case the low momentum part of
sterile neutrino distribution function is given
by~\cite{Gorbunov:2008ui,Tkachev, decays, Boyanovsky} \be f (p) = \frac83
\frac{\mpl^* \Gamma}{M^2} \l\frac{T_{0,eff}}{p}\r^{1/2} \int\limits_{0}^\infty
z^{3/2} f_{th}(z)\,dz =\frac{\zeta(5/2)}{4\pi^{5/2}} \ \frac{\mpl^*
\Gamma}{M^2}\ \l\frac{T_{0,eff}}{p}\r^{1/2} \; .
\label{aug14-1}
\ee
Here $\mpl^* \equiv \mpl\sqrt{90/(8\pi^3 g_*)}$, $T_{0, eff}= \l
\frac{g_{*,0}}{g_*} \r^{1/3} T_0 $, where $g_{*}$ and $g_{*,0}$ are the
effective numbers of degrees of freedom at decay and present epoch,
respectively, and $f_{th}$ is the thermal distribution function of decaying
particles. We assume here that the latter are scalars.  Hereafter $f(p)$ denotes
the primordial distribution function redshifted to the present epoch.  The total
present number density is~\cite{Gorbunov:2008ui}
$$
n_0 = \frac{3\zeta(5)}{4\pi} T_{0,eff}^3 \ \frac{\mpl^* \Gamma}{M^2} \; .
$$
Requiring that sterile neutrinos make all of dark matter,
$n_0 m = \Omega_\text{DM} \rho_c$, one finds the only relevant combination
of parameters of decaying particles,
 $$
\frac{\Gamma}{M^2} = \frac{4\pi}{3 \zeta(5)} \ 
\frac{\Omega_\text{DM} \, \rho_c}{m \, \mpl^* T_{0,eff}^3} \; .
$$ The distribution \eqref{aug14-1} is formally unbounded from above. In reality
this means that at low momenta, the distribution function takes the Pauli
blocking value, $f= 2/(2\pi)^3$. In this situation one in principle is still
able to obtain mass limits stronger than \eqref{aug14-2} by invoking the
following statistical argument~\cite{Madsen}. One requires that in the early
Universe, a certain fraction $\nu$ of dark matter particles are sufficiently
densely packed in phase space so that these particles are able to form
subsequently dark matter halos of high $Q$.  In other words, the value of the
distribution function of this fraction of particles should obey the constraint
\be f(p) > \frac Q {3^{3/2}\, m^4} \; ,
\label{aug14-3}
\ee where the observed value of $Q$ is estimated as in \eqref{not-required}. The
fraction $\nu$ should not be smaller than the fraction of dark matter residing
in dSph's, which we estimated in~\cite{Gorbunov:2008ui} as
\[
\nu \sim 10^{-5} \; .
\]
Given very weak dependence on $\nu$ in the limits we obtain in what follows,
the precise number is unimportant for our purposes.

We continue the discussion of thermal creation by noticing that the fraction
$\nu$ of most densely packed particles is related to the maximum momentum
$p_\nu$ of these particles in an obvious way,
\[
\int^{p_\nu}_0 f(p)\,4\pi p^2 dp = \nu n_0 \; .
\]
Making use of \eqref{aug14-1} we obtain
\[
\frac{p_f}{T_{0,eff}} = \l \frac{15\,\sqrt\pi\, \zeta(5)}{8\, \zeta(5/2)} \ \nu
\r^{2/5} \; .
\]
Substituting this value of momentum back into \eqref{aug14-1} we find
\begin{align*}
 f(p_\nu) &= \frac{\Omega_\text{DM} \, \rho_c}{m \, T_{0,eff}^3}
\frac1{3 \pi^{8/5}} \l\frac{\zeta(5/2)}{\zeta(5)}\r^{6/5}
\l\frac8{15}\r^{1/5} \ \nu^{-1/5} \\ &= 1.1\cdot 10^{-2} 
\l \frac{\Omega_\text{DM}}{0.2} \r
\l\frac{g_{*}}{106.75}\r
\l\frac{1\,\text{keV}}{m}\r \l\frac{10^{-5}}{\nu}\r^{1/5}\; .
\end{align*}
It is this value that should obey the constraint \eqref{aug14-3}.
Hence, we obtain the lower bound on the sterile neutrino mass,
\be
 m > 
0.88\,\text{keV}\ \l \frac{0.2}{\Omega_\text{DM}} \r^{1/3}
\l\frac{106.75}{g_*}\r^{1/3}
\l \frac{q}{5\cdot 10^{-3}} \r^{1/3}
\l\frac{\nu}{10^{-5}}\r^{1/15} \; .
\label{aug14-10}
\ee This bound is in fact slightly weaker than the Pauli blocking bound
\eqref{aug14-2} for our values of parameters. As we noticed above, in this
situation one should use the bound \eqref{aug14-2} instead.

The bound \eqref{aug14-10} is somewhat different from the bound obtained in
Ref.~\cite{Boyanovsky}, since we use the statistical approach rather than the
approach of Ref.~\cite{Hogan}. Notice that the dark matter is warm, i.e., the
bound \eqref{aug14-2} is nearly saturated if the parameters $M$ and $\Gamma$ are
such that
\[
M \frac{M}{\Gamma} \sim 10^{19}\,\text{GeV}\ \l \frac{0.2}{\Omega_\text{DM}}
\r \l\frac{106.75}{g_*}\r^{3/2}\;.
\]
This is the case if either the decaying particles are very heavy or their
decay rate into sterile neutrinos is very small, or both.

Finally, let us consider the production of sterile neutrinos in decays of
non-relativistic particles whose number in comoving volume
has been frozen out. 
In that case the momentum of sterile neutrinos at production equals
$p_* = M/2$, and then the momentum gets redshifted to
\be
p = p_*\,\frac{a(t)}{a (t_0)} = p_* \frac{T_{0 , eff}}{T}\; ,
\label{aug14-21}
\ee
where $t$ is the time at decay and $t_0$ is the present time.
Hence, the distribution function of sterile neutrinos is obtained from
the relation
\[
f(p)\, d^3 p = n_0 \, e^{-\Gamma_{tot} \cdot t}\, \Gamma_{tot}\, dt \; .
\]
Notice that we normalized the distribution function to the present number
density of sterile neutrinos, assuming that the abundance of decaying particles
is just right to produce them.  Therefore, this formula contains the total width
$\Gamma_{tot}$ only.  Making use of \eqref{aug14-21} we obtain 
\be f(p) =
n_0\,e^{-\Gamma_{tot}\cdot t} \ \frac{\Gamma_{tot}}{H(t)}\ \frac1{4\pi p^3 }\; \; .
\label{aug14-12}
\ee Now, at radiation dominated epoch\footnote{One can check that the energy
density of decaying non-relativistic particles never dominates in the case we
consider here.}  $t= \mpl^*/(2T^2)$, so that from \eqref{aug14-21} we obtain \be
t = \frac{\mpl^*}{2 T_{0 , eff}^2} \l \frac{p}{p_*} \r^2 \; , \;\;\;\;\; H(t) =
\frac{1}{2t} \; .
\label{aug14-20}
\ee
Hence, the distribution function \eqref{aug14-12} behaves as $1/p$ at low
momenta, and we again have to employ the statistical argument. Time $t_\nu$ by
which the fraction $\nu$ of sterile neutrinos is produced, is determined by
\[
\nu = 1-e^{-\Gamma_{tot}\cdot t_\nu} = \Gamma_{tot} \, t_\nu \;.
\]
The corresponding momentum $p_\nu$ is found from \eqref{aug14-20}. Inserting it
into \eqref{aug14-12} and requiring that sterile neutrinos make all of dark
matter, we find
\begin{align*}
 f(p_\nu) &= \frac{\sqrt2}{\pi} \ \frac{\Omega_\text{DM} \, \rho_c}{m \,
T_{0,eff}^3} \l\frac{\mpl^* \Gamma_{tot}}{M^2}\r^{3/2} \nu^{-1/2}\\ &= 2.4 \;
\l\frac{\mpl^*\Gamma_{tot}}{M^2}\r^{3/2}\l \frac{\Omega_\text{DM}}{0.2} \r
\l\frac{g_{*}}{106.75}\r \l\frac{1\,\text{keV}}{m}\r
\l\frac{10^{-5}}{\nu}\r^{1/2}\;.
\end{align*}
We again make use of \eqref{aug14-3} for this value of the distribution
function, and obtain finally the bound 
\be m > 145\,\text{eV} \
\l\frac{M^2}{\mpl^* \Gamma_{tot}}\r^{1/2} \cdot \l \frac{0.2}{\Omega_\text{DM}}
\r^{1/3} \l\frac{106.75}{g_*}\r^{1/3} \l \frac{q}{5 \cdot 10^{-3}} \r^{1/3}
\l\frac{\nu}{10^{-5}}\r^{1/6}\; .
\label{aug14-30}
\ee
This bound should be used whenever it supersedes the bound \eqref{aug14-2}.

Unlike the limit \eqref{aug14-10}, the bound \eqref{aug14-30} depends
on parameters $M$ and $\Gamma_{tot}$ characterizing the decaying
particles.  This is because unlike in the thermal production case, we
now have (implicitly) one more free parameter, the number density of
decaying particles at their freeze-out.  Interestingly, for heavy
enough decaying particles and/or long enough lifetime of these
particles, the right hand side of \eqref{aug14-30} may be well above
the ``canonical'' keV range. This implies that the decay mechanism we
discuss here is capable of producing {\it warm} dark matter composed
of {\it heavy} particles. As an example, for $M \simeq 10^{14}$~GeV,
$\Gamma_{tot} = \frac{y^2}{8\pi} M$ and $y \simeq 10^{-12}$, sterile
neutrino of mass in TeV range would be warm.

It is worth noting that the bounds \eqref{aug14-10} and
\eqref{aug14-30} apply not only to sterile neutrinos but to any
fermionic dark matter candidates produced in similar decay processes.

\vspace{0.5cm}

After this work has been completed, we received a draft of the paper~\cite{Boyarsky:neutrino},
where similar issues have been considered. Our results are consistent
with the results of Ref.~\cite{Boyarsky:neutrino} wherever they overlap.

\vspace{0.5cm}

{\bf Acknowledgments.}  We are indebted to F.~Bezrukov, A.~Boyarsky, S.~Demidov,
V.~Lukash, O.~Ruchayskiy, M.~Shaposhnikov and I.~Tkachev for useful discussions.
This work was supported in part by the grants of the President of the Russian
Federation NS-1616.2008.2 and MK-1957.2008.2 (DG), by the RFBR grant
08-02-00473-a and by the Russian Science Support Foundation (DG).

\end{document}